\documentclass{article}

\usepackage[top=1in, bottom=1in, left=1.25in, right=1.25in]{geometry}
\usepackage{authblk}

\usepackage{graphicx,graphics,wrapfig,rotating}
\usepackage{epsfig}
\usepackage{dcolumn}
\usepackage{bm,fancybox}
\usepackage{amsmath,amssymb}
\usepackage{times,euscript,oldgerm}
\usepackage[english]{babel}
\usepackage{multirow}
\usepackage{psfrag}
\usepackage{color}
\usepackage{longtable}
\usepackage[absolute,overlay]{textpos}
\usepackage{tikz}
\usetikzlibrary{decorations.pathmorphing,shadows,shapes}

\usepackage{hyperref}

\newcommand{\Yb}{\ensuremath{^{171}\mathrm{Yb}^+~}}

\newcommand{\ket}[1]{\ensuremath{\left|#1\right\rangle}}

\newcommand{\bra}[1]{\langle{#1}|}

\begin{document}

\title{Fermion-antifermion scattering via boson exchange in a trapped ion}

\author{Xiang Zhang$^{1,2}$, Kuan Zhang$^1$, Yangchao Shen$^1$, Jingning Zhang$^1$, Man-Hong Yung$^{3,1}$, \\Jorge Casanova$^4$, Julen S. Pedernales$^5$, Lucas Lamata$^5$, Enrique Solano$^{5,6}$, Kihwan Kim$^1$}

\affil{\small{
$^1$ Center for Quantum Information, Institute for Interdisciplinary Information Sciences, Tsinghua University, Beijing 100084, P. R. China \\
$^2$ Department of Physics, Renmin University of China, Beijing 100073, P. R. China\\
$^3$ Department of Physics, South University of Science and Technology of China, Shenzhen 518055, P. R. China\\
$^4$ Institut f\"ur Theoretische Physik and IQST, Albert-Einstein-Allee 11, Universit\"at Ulm, D-89069 Ulm, Germany\\
$^5$ Department of Physical Chemistry, University of the Basque Country UPV/EHU, Apartado 644, 48080 Bilbao, Spain \\
$^6$ IKERBASQUE, Basque Foundation for Science, Maria Diaz de Haro 3, 48013 Bilbao, Spain

}}
\date{\today }

\maketitle

\begin{abstract}
Quantum field theories describe a wide variety of fundamental phenomena in physics. However, their study often involves cumbersome numerical simulations. Quantum simulators, on the other hand, may outperform classical computational capacities due to their potential scalability. Here, we report an experimental realization of a quantum simulation of fermion-antifermion scattering mediated by bosonic modes, using a multilevel trapped ion, which is a simplified model of fermion scattering in both perturbative and nonperturbative quantum electrodynamics. The simulated model exhibits prototypical features in quantum field theory including particle pair creation and annihilation, as well as self-energy interactions. These are experimentally observed by manipulating four internal levels of a $^{171}\mathrm{Yb}^{+}$ trapped ion, where we encode the fermionic modes, and two motional degrees of freedom that simulate the bosonic modes. Our experiment establishes an avenue towards the efficient implementation of fermionic and bosonic  quantum field modes, which may prove useful in scalable studies of quantum field theories in perturbative and nonperturbative regimes.
\end{abstract}

\vspace{1cm}

Quantum simulators are devices designed to predict the properties of physical models associated with target quantum systems~\cite{Feynman82,Lloyd96}. Their intrinsic physical behaviors are fully governed by the laws of quantum mechanics, making it possible to efficiently study complex quantum systems that cannot be solved by classical computers~\cite{CiracZoller12,Nori14}. Trapped ions and superconducting circuits have proved to be promising for experimentally simulating a variety of paradigmatic quantum models such as various spin models~\cite{Leibfried02,Schaetz08,Kim10,Lanyon11,Barends16}, relativistic Dirac equations~\cite{Lamata07,Gerritsma10,CasanovaKlein,InnsbruckKlein}, embedding quantum simulators~\cite{PhysRevX.1.021018,ncomms8917,WhiteEmbeddings,PanEmbeddings,DiCandiaEmbeddings} and fermionic models~\cite{PhysRevLett.114.070502,ncomms8654}. More recently, a digital quantum simulation of a fermionic lattice gauge theory has been performed in trapped ions~\cite{Blatt16}. However, it would be desirable to realize a quantum simulator that involves interacting fermionic and bosonic fields as described by quantum field theories~(QFT)~\cite{Peskin}. In this sense, fermionic modes could be mapped in the ion internal levels, while bosonic modes could be naturally encoded in the motional degrees of freedom. 

Here, we report the first experimental quantum simulation of interacting fermionic and bosonic quantum field modes, where fermions are encoded in four internal levels of an Ytterbium ion and the bosonic modes in two motional modes, following the proposal by Casanova et al.~\cite{Casanova11b}. Therefore, this analog quantum simulation constitutes a milestone towards a digital-analog quantum simulator~\cite{PhysRevLett.114.070502,Casanova12,Mezzacapo14,Arrazola16,Lamata16} of perturbative and nonperturbative quantum field theories. In this sense, a remarkable feature of our experiment is that it contains all orders in perturbation theory, which is equivalent to all Feynman diagrams for a finite number of fermionic and bosonic modes. Moreover, our approach can be scaled up by adding progressively more ions allowing the codification of additional fermionic and bosonic field modes, which may lead to full quantum simulations of quantum field theories such as quantum electrodynamics (QED)~\cite{Peskin}.

\section*{Results}

The most common classical way to analyze QFTs is via a Dyson series expansion in perturbation theory and Feynman diagrams~\cite{Peskin}. If we consider larger coupling parameters, standard perturbative methods become cumbersome for a finite-mode Dyson expansion, mainly because only a reduced number of perturbative Feynman diagrams can be calculated. On the other hand, a trapped-ion quantum simulator with its high degree of quantum control~\cite{Leibfried03} could overcome these limitations and simulate QFTs more efficiently than classical computers~\cite{Preskill}. Based on the proposal of Ref.~\cite{Casanova11b}, our experimental quantum simulation of finite-number interacting quantized field modes includes all terms of the Dyson expansion. We experimentally implement a fundamental QFT model in a single trapped-ion considering i) one fermionic and one antifermionic field modes, ii) one or two bosonic field modes, which already reveals interesting QFT features such as self-interactions, particle creation and annihilation, perturbative and nonperturbative regimes. The general Hamiltonian involving the continuum of fermionic and bosonic fields reads
\begin{eqnarray}\nonumber
H &= \int dp\, \omega (b^\dag_pb_p+d^\dag_pd_p)+\int dk \, \omega_{k}a_{k}^{\dagger}a_{k}\\
&+g\int dx \psi^{\dagger}(x)\psi(x)A(x),
\end{eqnarray} 
where $b_p$ and $d_p$ are fermionic and antifermionic annihilation operators, respectively, while $a_k$ are the bosonic annihilation operators. Here, $\omega$ ($\omega_k$) is the fermion and antifermion free energy (boson free energy), while $\psi(x)$ denotes the fermionic and $A(x)$ the bosonic fields~\cite{Casanova11b}.

\begin{figure}
\centering
\includegraphics[width=0.5\textwidth]{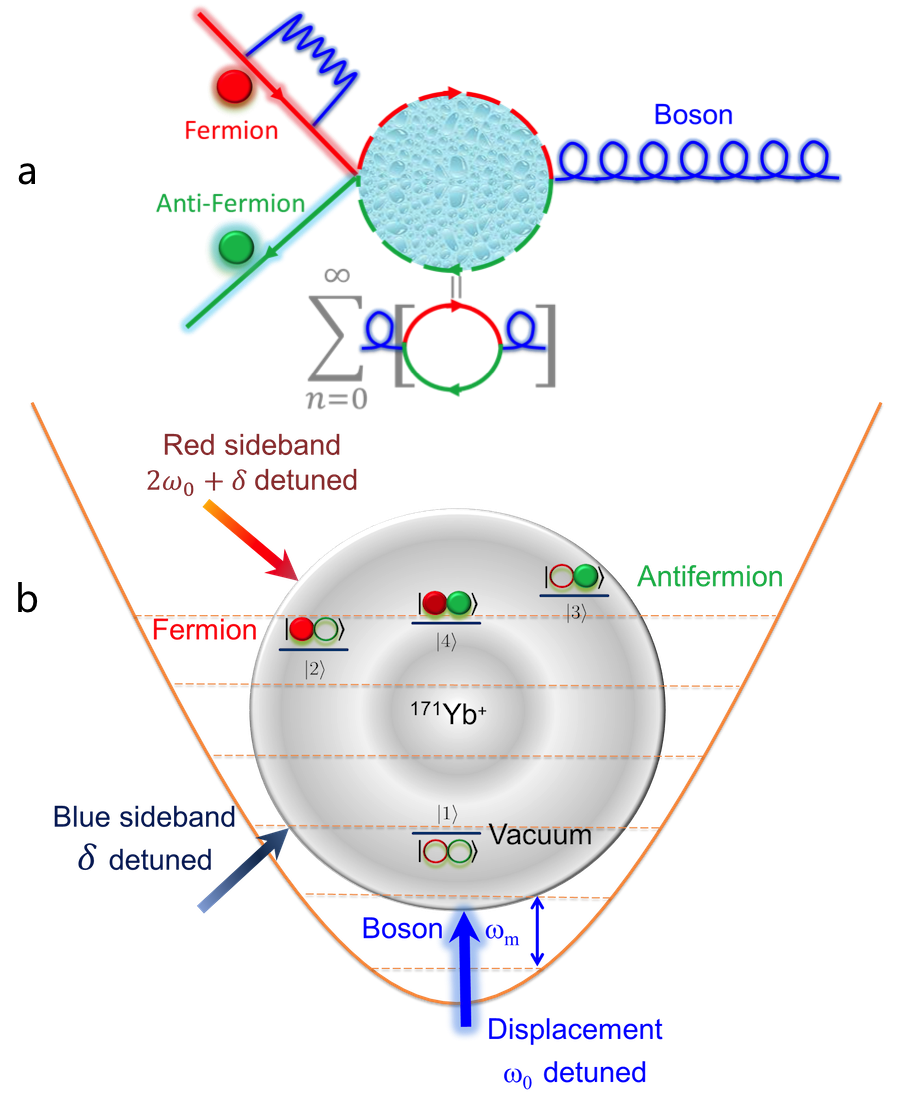}
\caption{(a) Diagram of interactions between fermion, antifermion, and bosons. \label{fig:FeynmanDiagram} (b) Diagram of the encoding of a QFT experiment in a trapped ion.\label{fig:QFTExperiment}}
\end{figure}

As a stepped experimental demonstration, we first consider the simplest situation with only one bosonic mode, which can be represented as a single vibrational mode of the ion along the X or Y direction. The fermion and antifermion modes are considered as two comoving modes describing incoming Gaussian wave packets which are centered in the average momentum with distant average initial positions~\cite{Casanova11b}. These modes describe self-interaction dressed states by emission and absorption of virtual bosons. They also represent the lowest-order in perturbation theory of the scattering of the incoming wave packets that will collide in a certain region of spacetime. The pair creation and annihilation is local and takes place only when the two wave packets of the fermion and antifermion overlap, namely, when the particles scatter. A diagram of these interactions, in the spirit of a Feynman diagram, is shown in Fig. \ref{fig:FeynmanDiagram}. Note that the loop of this diagram includes all terms in a finite-mode Dyson expansion, which is different from the standard perturbative approach with only a reduced number of Feynman diagrams. By considering slow massive bosons, as described in Ref.~\cite{Casanova11b}, the interaction Hamiltonian we would like to realize turns into
\begin{eqnarray}
H & = & g_1 e^{-i \omega_0 t}(b^\dagger b a_0 + d d^\dagger a_0) + g_2 e^{-(t-T/2)^2/(2\sigma_t^2)}\nonumber\\
&& \times (e^{i \delta t}b^\dagger d^\dagger a_0+e^{-i (2\omega_0+\delta) t}d\, b\, a_0) + {\rm h.c.},
\end{eqnarray}
where $\delta=\omega_f + \omega_{\overline{f}}-\omega_0$. Here, $\omega_f$, $\omega_{\overline{f}}$, and $\omega_0$ represent the energy of the fermionic field mode $b$, the antifermionic field mode $d$, and the bosonic field mode $a_0$, respectively. The ratio $g_2/g_1$ gives the relative strength between pair creation and self-interaction. $T$ is the total time of the pair-creation process while $\sigma_t$ is the temporal interval of the interaction region.

Applying a Jordan-Wigner mapping~\cite{Casanova11b} from fermionic modes to four internal levels of a single \Yb ion, 
\begin{eqnarray}
& b^{\dagger} = I\otimes\sigma^{+},b=I\otimes\sigma^{-},\nonumber\\
& d^{\dagger} = \sigma^{+}\otimes\sigma_{z},d=\sigma^{-}\otimes\sigma_{z},
\end{eqnarray}
the Hamiltonian becomes
\begin{eqnarray}
\label{Hi}
H_I & = & g_1(\ket 1\bra 1+2 \ket 2\bra 2 + \ket 4\bra 4)a_{0}e^{-i \omega_{0}t}\nonumber\\
&-& g(t)(\ket 1\bra 4 a_{0}^{\dagger}e^{-i \delta t}+\ket 1\bra 4 a_{0} e^{-i (2\omega_{0}+\delta)t}) + {\rm h.c.},
\end{eqnarray}
where the associated time-dependent coupling strength
\begin{equation}
g(t) = g_{2}e^{-(t-T/2)^{2}/(2\sigma_{t}^{2})} .
\end{equation}
The vacuum state and  fermionic states are represented by
\begin{equation}
\ket 1 = \ket{\mathrm{vacuum}}, \ket 2 = \ket f, \ket 3 = -\ket{\overline{f}}, \ket 4 = -\ket{f,\overline{f}},
\end{equation}
while $\ket{f,\overline{f},n}$ denotes the state containing one fermion, one antifermion, and $n$ bosons, respectively. The experimental diagram for this implementation is shown in Fig. \ref{fig:QFTExperiment}.

\begin{figure}
\centerline{
  \resizebox{0.5\textwidth}{!}{
    \begin{tikzpicture}[
      scale=1.5,
      level/.style={thick},
      virtual/.style={thick,densely dashed},
      trans/.style={thick,shorten >=1pt,shorten <=3pt,>=stealth},
    ]
    \node at (-3,0.6) {$^2\mathrm{S}_{1/2}$};
    \draw[level] (-0.5,0) node[coordinate,left] (s1l) {} -- node[midway,fill=white] {$\ket{1}$} +(1,0) node[coordinate,midway] (s1) {} node[coordinate,right] (s1r) {};
    \draw[virtual] (-0.5,0.4) node[coordinate,left] (s11l) {} -- +(1,0) node[coordinate,midway] (s11) {} node[coordinate,right] (s11r) {};
    \draw[>=stealth,<->,gray,dashed] (s1l) -- node[left] {$\omega$} (s11l);
    \draw[level] (-0.5,1) node[coordinate,left] (s3l) {} -- node[midway,fill=white] {$\ket{4}$}  +(1,0) node[coordinate,midway] (s3) {} node[coordinate,right] (s3r) {};
    \draw[virtual] (-0.5,1.4) node[coordinate,left] (s31l) {} -- +(1,0) node[coordinate,midway] (s31) {} node[coordinate,right] (s31r) {};
    \draw[>=stealth,<->,gray,dashed] (s3l) -- node[left] {$\omega$} (s31l);
    \draw[level] (-2,0.8) node[coordinate,left] (s2l) {} -- node[midway,fill=white] {$\ket{2}$} +(1,0) node[coordinate,midway] (s2) {} node[coordinate,right] (s2r) {};
    \draw[virtual] (-2,1.2) node[coordinate,left] (s21l) {} -- +(1,0) node[coordinate,midway] (s21) {} node[coordinate,right] (s21r) {};
    \draw[>=stealth,<->,gray,dashed] (s2l) -- node[left] {$\omega$} (s21l);
    \draw[level] (1,1.2) -- +(1,0) node[coordinate,midway] (s4) {};
    \node at (-3,5.8) {$^2\mathrm{P}_{1/2}$};
    \node at (-3,4) {$\Delta\approx 12$ THz};
    \node at (-3,3) {$\omega = \omega_m - \omega_0$};
    \draw[level] (-0.5,5.5) -- +(1,0) node[coordinate,midway] (p1) {} node[coordinate,right] (p1r) {};
    \draw[virtual] (-0.5,4.8) node[coordinate,left] (p1dl) {} -- +(1,0) node[coordinate,midway] (p1d) {} node[coordinate,right] (p1dr) {};
    \draw[>=stealth,<->,gray,dashed] (p1r) -- (p1dr) node[midway,right] {$\Delta$};
    \draw[level] (-0.5,6) node[coordinate,left] (p3l) {} -- +(1,0) node[coordinate,midway] (p3) {};
    \draw[virtual] (-0.5,5.3) node[coordinate,left] (p3dl) {} -- +(1,0) node[coordinate,midway] (p3d) {} node[coordinate,right] (p3dr) {};
    \draw[>=stealth,<->,gray,dashed] (p3l) -- (p3dl)  node[midway,left] {$\Delta$};
    \draw[level] (-2,5.9) -- +(1,0) node[coordinate,midway] (p2) {};
    \draw[level] (1,6.1) -- +(1,0) node[coordinate,midway] (p4) {};
    \draw[virtual] (1,5.4) node[coordinate,left] (p4dl) {} -- +(1,0) node[coordinate,midway] (p4d) {} node[coordinate,right] (p4dr) {};
    \draw[>=stealth,<->,gray,dashed] (p4) -- node[midway,fill=white] {$\Delta$} (p4d);
    \draw[trans,<->,blue] (s1r) -- (p4dr);
    \draw[trans,<->,blue] (s11l) -- (p4dr);
    \draw[trans,<->,blue] (s3r) -- (p4dl);
    \draw[trans,<->,blue] (s31l) -- (p4dl);
    \draw[trans,<->,blue] (s2r) -- (p1dr);
    \draw[trans,<->,blue] (s21l) -- (p1dr);
    \draw[trans,<->,blue] (s2r) -- (p3dl);
    \draw[trans,<->,blue] (s21l) -- (p3dl);
    \end{tikzpicture}
    }
}
\caption{Diagram of the displacement operator with $\sigma_+$-polarized Raman beams. \label{fig:Displacement}}
\end{figure}
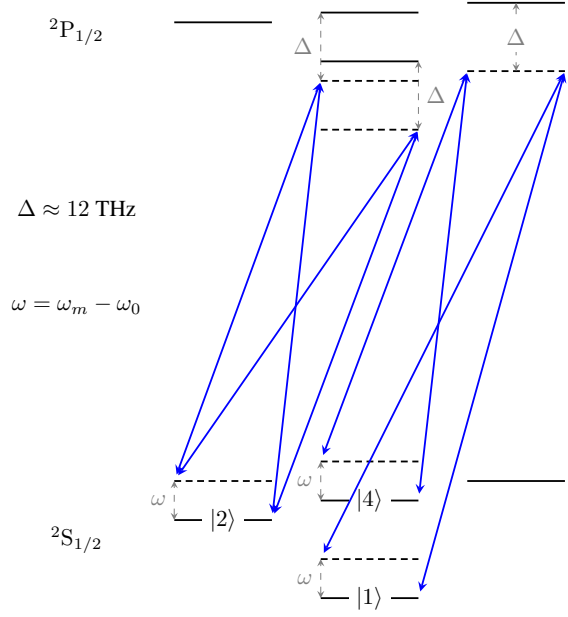

The Hamiltonian in Eq.~(\ref{Hi}) can be implemented with the following trapped-ion operations:
\begin{itemize}
\item $(\ket{1}\bra{1}+2\ket{2}\bra{2}+\ket{4}\bra{4})a_{0}e^{-i \omega_{0}t}$: $\omega_0$ detuned displacement operators with $\sigma_+$-polarized Raman beams, as shown in Fig. \ref{fig:Displacement}.
\item $\ket{1}\bra{4}a_0^{\dagger}e^{-i \delta t}$:
$\delta$ detuned red sideband transition between $\ket{1}\leftrightarrow\ket{4}$.
\item $\ket{1}\bra{4}a_0e^{-i (2\omega_0+\delta) t}$:
$2\omega_0+\delta$ detuned blue sideband transition between $\ket{1}\leftrightarrow\ket{4}$.
\end{itemize}

All optical transitions between $^2\mathrm{S}_{1/2}\leftrightarrow ^2\mathrm{P}_{1/2}$ states have the same transition strength factor with respect to absolute values. For this experimental case, the $\sigma_+$-polarized Raman beams with frequency difference $\omega_m-\omega_0$ give exactly the desired displacement Hamiltonian $(\ket{1}\bra{1}+2\ket{2}\bra{2}+\ket{4}\bra{4})a_{0}e^{-\imath\omega_{0}t}$. In principle, we can also implement other kinds of displacement Hamiltonians by applying additional $\sigma_-$ and $\pi$-polarized Raman beams with specific ratios.

This experiment is realized with stimulated Raman transitions implemented by a $375$ nm ``MiraHP'' mode-locked pulse laser. We first cool down vibrational modes and initialize the ion in the vacuum state $\ket{1,n=0}$ by resolved sideband cooling \cite{Monroe95a}. Then, a phonon displacement Raman beam component for all internal levels, a detuned blue sideband Raman beam component and a detuned red sideband Raman beam component (both for the clock transition) are applied simultaneously to realize this Hamiltonian in our trapped \Yb ion system. All these Raman laser beams introduce state-dependent forces and push the ion along $\pm\Delta k$ direction for certain internal states. After the evolution process, we measure the final distribution of phonon number states by applying blue and red sideband transitions, while fitting the signals through the maximum likelihood method with parameters of the Fock state populations. For the measurement step, we observe the time evolution of blue and red sideband transitions up to 250 $\mu$s with 1 $\mu$s step by averaging over $200$ repetitions in each step.

For the average boson (phonon) number measurement, we first use optical pumping to trace out internal states and then make a phonon number fitting measurement with blue sideband time sweep. The populations of fermion mode or fermion-antifermion pair at bosonic mode $\ket{n=0}$ are measured based on the population measurement of state $\ket{1,0}$.  In order to measure the population of $\ket{1,0}$, we perform the following experimental steps.
\begin{itemize}
\item Collapse bright states population to dark state $\ket 1$ by standard detection process
\item A ``uniform red sideband'' transition to transfer states $\ket{1,n\neq 0}$
to bright state $\ket{3,n-1}$ with the same Rabi frequency
\item Carrier transition to swap bright state $\ket{3}$ and dark state $\ket{1}$
\item Standard detection process to measure bright states population, which should be equal to the original population of $\ket{1,0}$
\end{itemize}

\begin{figure}
\centering
\includegraphics[width=\linewidth]{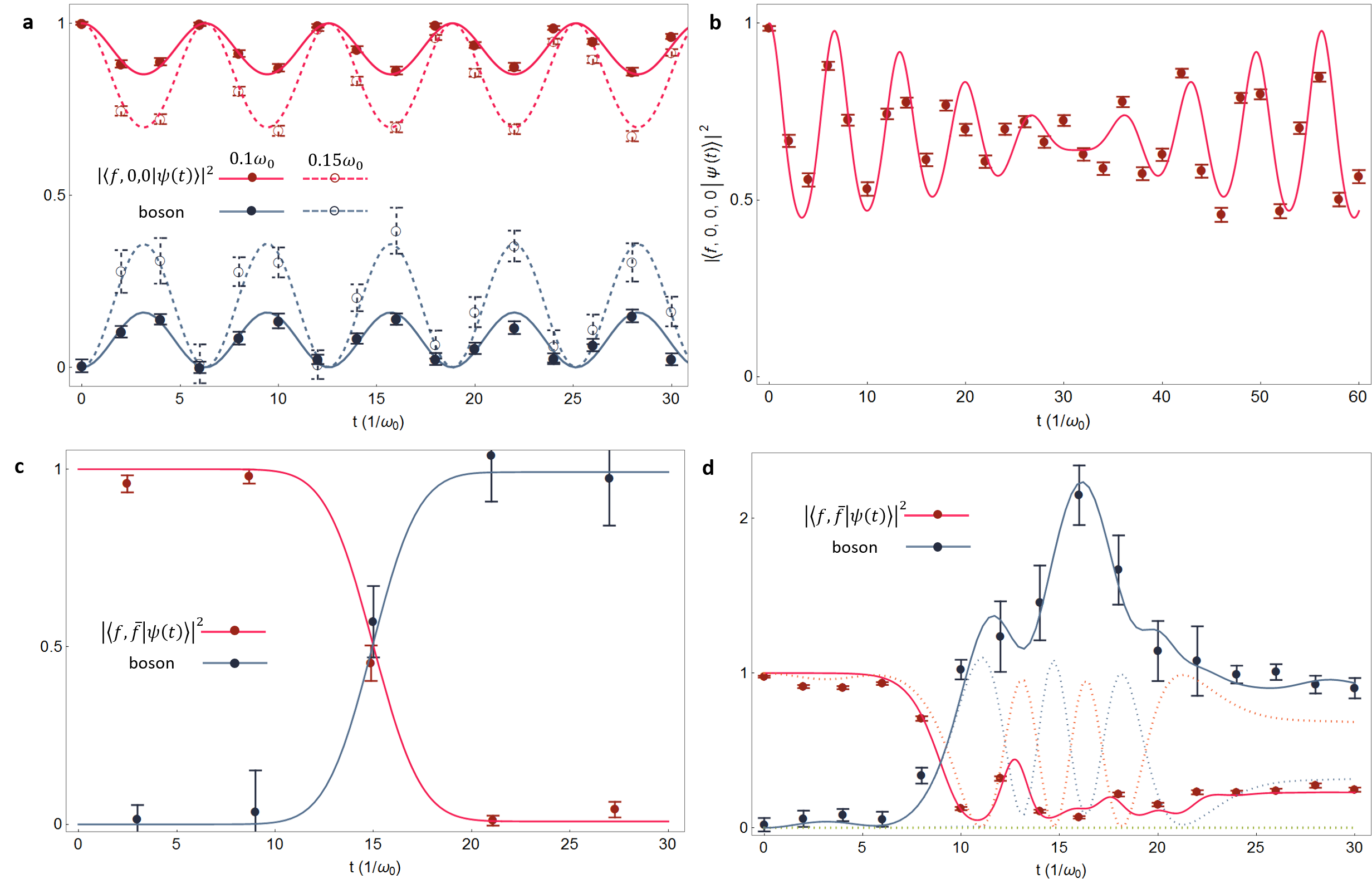}
\caption{Trapped-ion experimental results of the quantum simulation of fermion-antifermion scattering mediated by boson exchange in quantum field theory. Dots are experimental data and lines are numerical simulation curves. (a) Self-interaction process, with $g_1=0.1\omega_0,0.15\omega_0$. \label{fig:SelfInteraction} (b) Self-interaction process for two bosonic modes, with $\omega_1=\omega_0$ and $\omega_2=0.9\omega_0$. (c) Fermion and antifermion annihilation process. \label{fig:PairCreation} (d)~Nonperturbative interaction process. Note that solid lines are numerical simulation curves. Dotted lines are 6th order perturbation calculation with 50 iterations for each order by Feynman diagrams method. \label{fig:NonPerturbative}}
\end{figure}

We first realize the fermion self-interaction processes with parameters $$g_{2}=0,\sigma_{t}=3/\omega_{0}.$$ We choose the initial state to be one fermion state $\ket{f,0,0}$ with no bosons. Then the self-interacting dynamics is given by $\ket{f,0,n}\leftrightarrow\ket{f,0,n\pm 1}$. Full-fledged details of our experimental results are shown in Fig. \ref{fig:SelfInteraction}. We observe that the considered fermion emits and reabsorbs bosons at a period $2\pi/\omega_0$. The experimental results match well the theoretical simulation curve. This process is then extended to 2 bosonic modes by using both X and Y phonon modes of a single trapped ion. By performing additional sequential ``uniform red sideband'' transitions of X and Y modes, we calculate the population of the fermionic mode $\ket{f,0,0,0}=\ket{2,n_x=0,n_y=0}$ with the following set relation
\begin{equation}
\ket{2,n_x=0,n_y=0}=\ket{2,n_x= 0}-\ket{2,n_x= 0,n_y\neq 0} .
\end{equation}

Next, we realize the pair annihilation process with parameters $g_{1}=0.01\omega_{0},g_{2}=0.21\omega_{0},\sigma_{t}=3/\omega_{0}.$ We choose the initial state to be one fermion and one antifermion state $\ket{f,\overline{f},0}$ with no bosons. Then, the pair annihilation dynamics is given by $\ket{f,\overline{f},n}\leftrightarrow\ket{0,0,n\pm 1}$. Detailed experimental results are shown in Fig. \ref{fig:PairCreation}. We observe that the considered fermion and antifermion pair annihilates when the two modes overlap and enter the interaction region, a process giving rise to the population of the bosonic mode.

Finally, we realize the nonperturbative process with parameters $g_{1}=0.1\omega_{0},g_{2}=\omega_{0},\sigma_{t}=4/\omega_{0}.$ We also choose the initial state to be one fermion and one antifermion state $\ket{f,\overline{f},0}$ with no bosons. However, there is no simple analytic description for this strong coupling situation. The associated experimental data is shown in Fig. \ref{fig:NonPerturbative}. For large $g_2$ values ($g_2\geq \omega_0$), we run into the nonperturbative regime, where Feynman diagram techniques are not useful. When the two particles enter the interaction region, the theoretical curves calculated with Feynman diagrams strongly deviate from numerical simulation curves as well as from experimental data. We obtain that the number of created bosons is much larger due to the nonresonant terms present in the interaction. In this sense, the dynamics becomes more complex and strongly dependent on the specific coupling values.

\section*{Discussion}

In conclusion, this work can be considered as the first experimental quantum simulation of interacting fermionic and bosonic quantum field modes. Our approach can be scaled up by progressively incorporating more fermionic and bosonic field modes, which may lead to a full-fledged digital-analog quantum simulation of quantum field theories such as quantum electrodynamics (QED). In our current experimental system, an extension to multi phonon (bosonic) modes should be straightforward by loading a chain of ions and shining Raman laser beams at one end of the ion chain. This experiment opens an avenue that aims at outperforming the limitations of classical computers, with in principle scalable quantum simulations. In particular, we remark that already with 10 two-level ions and 5 phononic levels per ion, one could perform quantum simulations of interacting quantum field modes that are beyond the reach of classical computations, that is, a Hilbert space dimension of $10^{10}\sim2^{33}$, which would otherwise require a lengthy quantum algorithm with 33 qubits.

\section*{Acknowledgements}

This work was supported by the National Basic Research Program of China under Grants No. 2011CBA00300 (No. 2011CBA00301), the National Natural Science Foundation of China 11374178, 11405093, 11574002 and 11504197 as well as Spanish MINECO/FEDER FIS2015-69983-P, Ram\'on y Cajal Grant RYC-2012-11391, UPV/EHU UFI 11/55, Project EHUA14/04, a UPV/EHU PhD fellowship, the Alexander von Humboldt foundation, EU STREP EQUAM and ERC Synergy grant BioQ.


\end{document}